\documentclass{aastex}
\usepackage{spr-astr-addons}
\usepackage{url}

\RequirePackage{color}

\begin{document}

\title{Classifying GRB~170817A/GW170817 in a Fermi duration - hardness 
plane}
\slugcomment{Not to appear in Nonlearned J., 45.}
\shorttitle{Short article title}
\shortauthors{Autors et al.}

\author{I. Horv\'ath\altaffilmark{1}}
  \and \author{B.G. T\'oth\altaffilmark{1}}  
  \and \author{J. Hakkila\altaffilmark{2}}
  \and \author{L.V. T\'oth\altaffilmark{3,4}} 
  \and \author{L.G. Bal\'azs\altaffilmark{3,4}}  
    \and \author{I.I. R\'acz\altaffilmark{1,3}} 
  \and \author{S. Pint\'er\altaffilmark{3}}   
   \and \author{Z. Bagoly\altaffilmark{3}}


\altaffiltext{1}{National University of Public Service, Budapest, Hungary.}
\altaffiltext{2}{College of Charleston, Charleston, SC, USA.}
\altaffiltext{3}{E\"otv\"os  University, Budapest, Hungary.}
\altaffiltext{4}{MTA CSFK Konkoly Observatory, Budapest, Hungary.}

\begin{abstract}
GRB~170817A, associated with the LIGO-Virgo GW170817 neutron-star merger event,
lacks the short duration and hard spectrum
of a Short gamma-ray burst (GRB) expected from long-standing
classification models. Correctly identifying the class to which this burst
belongs requires comparison with other GRBs detected by the Fermi GBM.
The aim of our analysis is to classify Fermi GRBs and to test whether or not
GRB~170817A belongs -- as suggested -- to the Short GRB class.
The Fermi GBM catalog provides a large database with
many measured variables that can be used to explore
gamma-ray burst classification.
We use statistical techniques to
look for clustering in a sample of 1298 gamma-ray bursts described by
duration and spectral hardness.
Classification of the detected bursts
shows that GRB~170817A most likely belongs to
the Intermediate, rather than the Short GRB class.
We discuss this result in light of theoretical
neutron-star merger models and existing GRB classification
schemes. It appears that GRB classification schemes
may not yet be linked to appropriate theoretical models,
and that theoretical models may not yet
adequately account for known GRB class properties.
We conclude that GRB~170817A may not fit into a simple
phenomenological classification scheme.
\end{abstract}



\section{Introduction}\label{sec:intro}
One of the most exciting events in modern astrophysics has been the association of a
gravitational wave coalescence event with a gamma-ray burst.  On August 17,
2017, the LIGO and Virgo experiments \citep{PRL119} observed a chirp 
(GW170817) associated with the merger of two compact
objects in the mass range $1.17 - 1.60\ M_\odot$ with a combined mass of 
$2.74^{+0.004}_{-0.001}\ M_\odot$. The LIGO/Virgo chirp is thus consistent with
merging neutron stars. GRB~170817A triggered the Fermi  
Gamma-ray Burst Monitor (GBM) experiment
\citep{GCN21506} 1.7 s  after the gravitational wave event. 
Preliminary properties identified the GBM trigger 
\citep{GCN21528,2017GoldsteinVeres}  as having a
duration of $\approx2$ s and a 64 ms peak flux 
of $3.7 \pm 0.9\mathrm{\ ph/s/cm}^2$ ($10-1000$ keV).

At first glance, the association of the lower mass LIGO event with a Short GRB
seems to validate standard theoretical models based on known GRB classification.
Evidence from the 1980's suggested that two GRB classes existed on the basis
of duration \citep{maz81,nor84}. Subsequent observations provided by 
the Burst And Transient 
Source Experiment (BATSE) supported this division \citep{kou93,kos96} and also found
the Short GRBs to have harder spectra than the Long ones.
Compact objects are needed to produce large GRB luminosities and short timescales,
and the BATSE duration bimodality seemed to point to the existence of two 
distinctly different GRB populations. It was felt that although the timescale of massive 
star core-collapse was sufficient to explain Long GRBs, it was too long to explain Short GRBs.
As a result, theoretical models constraining progenitor compactness were merged with
observational evidence of clustered GRB properties to develop expectations of class properties.

For decades astronomers have sought clear
additional evidence that Short GRBs differ from Long GRBs other than
by duration and spectral hardness \citep{nor01,bal03,zha09,luli10,li16}.  
Some low-luminosity Long GRBs have been associated with
Type Ic supernovae (SN) \citep{hjor03,camp06,pian06,blan16}, supporting the
idea that the Long GRBs in general are related to deaths of massive stars
\citep{woo93,paczy98,wb06,blan16}.  For Short GRBs, the absence of SN
associations, the location of these events in metal-poor regions, and their
lower luminosities disfavor a massive star origin and point to compact binary
mergers \citep{paczy86,usov92,berger14}.
Observations supportive of these differences
have included GRB luminosities, different host galaxies and redshift distributions
\citep{berger14,levan16}, the metallicity of the
environment surrounding the GRB, different afterglow properties, etc. 
Thus, these supportive observations have led observers to believe that the
identification of a Short GRB with gravitational wave evidence of a neutron
star-neutron star merger would unambiguously demonstrate
the correctness of the standard model.

Despite the clear association of GRB~170817A with GW170817, the burst's duration, 
fluence, and soft spectrum allow for an uncomfortable ambiguity in its interpretation 
as a Short GRB. It is not clear that this object is either a Long or a Short GRB, as its properties 
straddle the boundary between the two classes. Most formal statistical classification 
techniques find at least one other class (the Intermediate class) occupying the space 
between the Long and Short GRB classes, although more statistical clusters have also 
been found. 

Using multi- and uni-variate statistical
analysis techniques,  \cite{muk98} and \cite{hor98} found evidence for a third
GRB class in data from the Third BATSE Catalog \citep{m6}.
The class is composed of GRBs having intermediate durations ($2\ \mathrm{s} \le \mathrm{T}_{90} \le 10$ s), 
intermediate fluences, and soft spectra characterized by soft hardness ratios. Many authors
\citep{hak00,bala01,rm02,hor02,hak03,bor04,hor06,chat07,zito15} have since
confirmed the existence of this Intermediate GRB class in the same database
using statistical techniques and/or data mining algorithms.  The Intermediate
class has also been found in the Beppo-SAX \citep{hor09} and Swift data
\citep{hor08,huja09,hor10,ht16}, even though Beppo-SAX had a smaller effective
area than BATSE, and Swift works in a different energy range.  
The properties of each class differ depending on instrumental characteristics 
of the experiment measuring them, the classification attributes being used, 
the classification techniques being applied, and the sample size. 
Through these analyses, class properties have been found to differ somewhat 
(the Short-Intermediate division typically occurs at longer durations for 
experiments other than BATSE), and the Short-Intermediate division has been 
found to be more robust than the Intermediate-Long division \citep{ugarte11}.

The $T_{\rm 90,BATSE} \approx 2$ s boundary separating Long and Short GRBs is not robust 
for a variety of reasons. 
First of all, it has been defined from GRBs observed by a single instrument (BATSE) 
having its own
surface area, spectral response, temporal resolution, and angular sensitivity. 
Second, it has been defined
from a bimodal interpretation of one specific dataset (defined in \cite{kou93}). 
Third, acceptance of this division has been
based partially on theoretical models rather than entirely on observational ones. 
Much of the GRB literature has incorrectly painted classification as a black-and-white 
division separating two distinctly different types of progenitors, 
whereas it is in reality a gray area occupied by observations of one distinct observational 
class (the Short class) separated in duration, fluence, and spectral hardness from at 
least one other (the Long class). As a result, the 2 s bimodal classification scheme 
should not be seen as being applicable to all GRB data, and especially not to data 
collected by GRB instruments other than BATSE. Instead, classification of GRBs 
collected by a specific instrument should be done independently, and interpretation 
should subsequently proceed based solely on observational evidence rather than on 
theoretical models. Classification is a meaningful way to regard data, but not theories. 

Published statistical clustering analyses have used a variety of different
variables in their classification approaches:
duration information  
\citep{hor98,bala01,rm02,hor02,hor08,huja09,hor09,zito15,
tarno15AA,tarno15ApSS,ht16,tarno16MNRAS,kulk17}, 
duration and hardness 
\citep{hor04,hor06,veres10,hor10,kb12,qin13,tsu14,sn15,rm16,yzj16,zycc16}, 
or more than two variables 
\citep{muk98,hak03,chat07,kk11,lu14,li16,2017arXiv170305532Modak,chat17}.

In this paper we only use duration and spectral
hardness to examine classification of the very interesting GRB~170817A, 
because we intended to fit this event into the scheme of prior analyses.
While trying to explain the observed peculiarities of the high energy 
emission of GRB~170817A, \citet{2017arXiv171007987B} speculate that it might 
represent a new (short) GRB sub-class. 

The paper is organized as follows. In Sect.~2 we present the cluster analysis of the Fermi data, 
Sect.~3 discusses the results, and Sect.~4 provides the summary and outlook.

\section{Cluster analysis with duration and hardness}

On September 12, 2017, the Fermi GBM 
Catalog\footnote{https://heasarc.gsfc.nasa.gov/W3Browse/fermi/fermigbrst.html} 
contained 2055 GRBs for which spectral fits were available, and listed more than 
300 parameters for those.
For our analysis we have chosen to use the duration ($\mathrm{T}_{90}$) and hardness
variables. The spectral hardnesses used here
have been kindly provided by Drs. Bhat and Veres (as also used in \citet{2017GoldsteinVeres}).

To improve the data quality we have excluded
GRBs having poorly-measured hardnesses.  
In order to retain as many GRBs in the sample as possible while still
minimizing the number of bursts with poor hardness measurements, we 
chose to exclude 79 GRBs having hardness uncertainties larger than 1.5 
times the hardness measurement. This leaves us
with 1298 GRBs for our analysis. 

Our analysis is made using the \textit{mclust} \citep{mbclust,mclust} 
package in the R environment \citep{rlang}.
The first step is to see whether there are any outliers
in the 1298-element dataset, using hardness and duration as our
classification variables. For this purpose the \textit{HDoutliers} package 
\citep{hdout} was used and no outliers were found.

\begin{figure}[h!]\begin{center}
 \resizebox{\hsize}{!}{\includegraphics[height=3.1cm,angle=0]{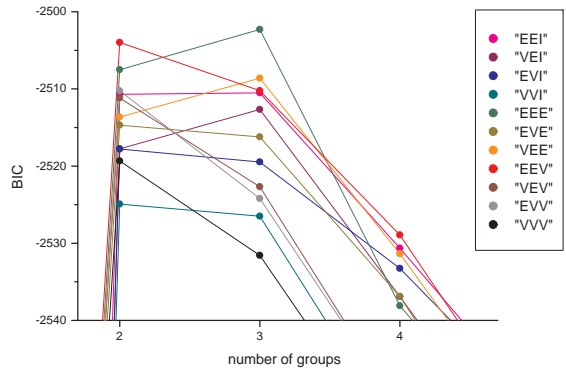} }
 
 \caption{\small{The optimal BIC value
 prefers three classes with EEE method (clusters having elliposidal distributions described by equal volumes, shapes and orientations).}}
  \label{fig:t90elo}
\end{center}
\end{figure}

We then proceeded to identify the optimal number of classes in the
hardness vs. duration parameter space using the Bayesian
Information Criterion (BIC). The BIC value was calculated using the \textit{Mclust()} 
function, initially
assuming $\mathrm{N}=1\ldots 
10$ groups for all the models available (Fig.~\ref{fig:t90elo}). 
The largest BIC value of $ -2502.29$ was obtained using the 
EEE model (assuming clusters having elliposidal distributions described by 
equal volumes, shapes and orientations) assuming three classes.  

\begin{figure}[h!]\begin{center}
 \resizebox{\hsize}{!}{\includegraphics[height=3.1cm,angle=0]{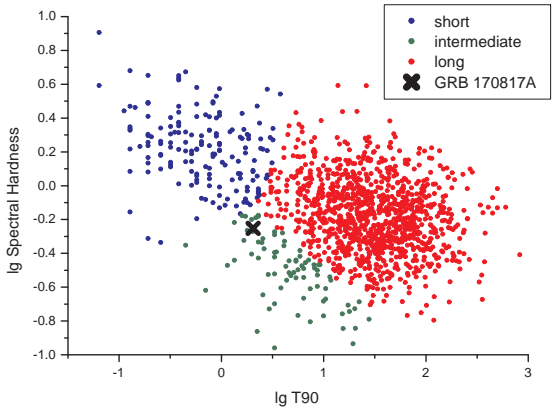} }
 \caption{\small{The $\log(\mathrm{T}_{90})$ - $\log(\mathrm{HR})$ distributions of the three classes.
 The Short GRB class is shown in blue, the Intermediate GRB class is in green, 
 and the Long GRB class is in red. GRB~170817A is clearly located in the 
 Intermediate group region.}}
  \label{fig:hrt90}
\end{center}
\end{figure}

The \textit{Mclust()} function also returns the probabilities $p_{Si}$, $
p_{Ii}$, and $p_{Li}$ that burst $i$ belongs to the Short (S), 
the Intermediate (I), or the Long (L) classes,
respectively. The assignment of each GRB according to the maximal $p_{ki}$
values gives the grouping plotted in Fig.~\ref{fig:hrt90}.
By summing these probabilities one gets $p_S = \sum_i p_{Si}=170.58$, $p_I = \sum_i p_{Ii}=130.21$, 
and $p_L = \sum_i p_{Li}=997.21$.

The Intermediate class can clearly be seen 
between the Long and Short GRB classes having the softest spectral hardness (Fig.~\ref{fig:hrt90}).
The general characteristics of the groups are similar to those 
found in BATSE \citep{hor98,muk98} and Swift \citep{hor08,veres10,hor10} data.

Based on its duration and hardness, GRB~170817A/ GW170817 
belongs to the Intermediate class ($p_{I,GW}=58.3\%$ against 
$p_{S,GW}=16.5\%$ and $p_{L,GW}=25.2\%$).
On the other hand, one can check the hypothesis
that GRB~170817A is a Short GRB.
Using the Fermi Short group parameters (see Table 1.)
and the position of the GRB~170817A in the 
duration - hardness plane, classification of 
GRB~170817A as a Short burst results in
a misclassification probability 94.0\%, which is near the
two sigma limit.

\begin{table}
\caption[]{Parameters for the best fitted three groups. $w$ is the weight of the group.}
		 \label{3g}
	 $$
		 \begin{array}{cccc}
		    \hline
		    \noalign{\smallskip}
	 & center (log T_{90}) &  log HR  & w \\
		    \noalign{\smallskip}
		    \hline
		    \noalign{\smallskip}
	short & -0.124 & 0.213   & 0.129		 \\
	intermediate & 0.906  & -0.404   & 0.062	\\
   long & 1.488  & -0.175   & 0.809	   \\
		    \noalign{\smallskip}
		    \hline
		 \end{array}
	 $$
\end{table} 

\section{Discussion}

The properties of the Intermediate GRB class are fuzzy because class properties overlap 
one another in the chosen parameter space. The choice of classification parameters is
generally based on what an instrument measures, rather than on an idealized
yet unknown parameter that might more clearly aid in class delineation.
The measured parameters and their corresponding measurement uncertainties
strongly depend on the instrumental characteristics and sampling biases. 
As a result, classification is instrument-dependent and should be
done separately for each orbital experiment.
This misapplication of GRB classification schemes has resulted in ambiguities that have, 
up until now, been ignored. 

\cite{kb12} classified Swift BAT GRBs and concluded that three
classes are needed to characterize the duration distribution,
whereas only two are are required to sufficiently describe the spectral 
hardness distribution. Swift's lower-energy spectral response is
likely responsible for weakening spectral hardness as a classification
attribute: high-energy emission is important in delineating BATSE classes.
The
Intermediate class identified by \cite{kb12} has durations of around $3-20$
seconds, which is in good agreement with \citep{ht16} who find the
Intermediate class durations to be in the $4-30$-second range.
We note here that \cite{kb12} assumed that the distribution is a 
combination of Gaussians, which is an assumption supported by \citet{ioka02}.

In addition to Swift, instrumental effects might also be responsible for
affecting Fermi classification results. \cite{qin13} analyzed 315
Fermi GRBs, studying the dependence of the duration distribution on energy and
on various instrumental and selection effects.  They have suggested that the
true durations of a GRB could be much longer than what is observed. 

Analyses of data from a variety of orbital high-energy satellites continue to
find evidence for three GRB classes over two. \cite{tsu09,tsu13,tsu14} have used data
from several orbital instruments, as well as X-ray and optical afterglow data,
to study GRB classes, and have found a third group with durations of $\mathrm{T}_{90,\mathrm{BAT}}
\approx 5$ s.  \cite{zito15} has analyzed the CGRO/BATSE and Swift/BAT GRB data
to find a very similar class structure to \citet{hor02}. 

Although most rigorous GRB classification studies find three classes in the
data, there have been exceptions.  In a recent publication,
\cite{tarno15ApSS} proposes that the division between Short and Long bursts
corresponds to a $\mathrm{T}_{90}$ of 3.4 seconds rather than two seconds. \cite{tarno15AA} analyzed the
Fermi GBM duration data of 1566 GRBs.  Although he found a third component in
the distribution, the significance was not convincing. This may be due to
methodology: the binned data were tested with a $\chi ^2$ fit rather than using a
maximum likehood method and unbinned data. This points to an additional
difficulty that has been found in applying statistical clustering techniques: the 
results depend on heuristic assumptions about the form that clustering takes and 
on the techniques most likely to extract this assumed clustering.

Although the three identified Fermi GBM classes of GRB overlap, 
GRB~170817A's prompt characteristics indicate that it is most likely an  
Intermediate GRB rather than a Short one  
(being a short GRB has only 6\%  probability).
Association of a LIGO chirp with an 
Intermediate GRB is itself inconvenient in that it requires modification of existing 
theoretical models as well as recognition of the existence of the Intermediate GRB class.

Our classification evidence suggests that GRB~170817A represents 
an Intermediate GRB rather than an outlier Short GRB. Accepting these
classification results along with the evidence from the gravitational wave
chirp suggests that Intermediate GRBs are associated with compact merger systems, 
If Intermediate GRBs are somehow
an extension of the Short GRB merger model, then a
more thorough characterization of Intermediate GRB afterglow
and host galaxy properties need to be made in order to determine
how they differ from those of traditional Short GRBs.
\cite{ugarte11} concluded that the
intermediate bursts are found to be less energetic and have
dimmer afterglows than long GRBs, especially when considering
the X-ray light curves, which are on average one order of
magnitude fainter than long bursts. There is a less significant
trend in the redshift distribution that places intermediate
bursts closer than long bursts. Except for this, intermediate
bursts show similar properties to long bursts. In particular,
they follow the $E_{peak}$ versus $E_{iso}$ correlation and have, on
average, positive spectral lags with a distribution similar
to that of long bursts. As for long GRBs, they normally
have an associated supernova, although some intermediate
bursts have been found to contain no supernova component.

It is still possible that GRB~170817A is an outlier Short GRB,
which would be more consistent with the statistical properties of
Short and Intermediate GRBs. Besides sharing more parameter 
space with the Longs than with the Shorts, Intermediates
extend a correlation found in Long GRBs where fainter Longs are
softer than brighter Longs ({\em e.g.} \cite{hak11}). 
Pulses of all GRB classes
exhibit fairly similar triple-peaked structures \citep{hak14,hak15, hak18}, 
suggesting that a common mechanism is
responsible for producing them regardless of their progenitors and/or
host galaxies. However, pulses produced by both Long and
Intermediate GRBs appear to have significantly longer durations than pulses
from Short GRBs \citep{hak14,hak18}.

The growing number of bursts detected by the Fermi 
GBM provides additional data on which GRB classification schemes can be
tested. GBM has a spectral energy response that is similar to, but broader,
than BATSE, and a surface area that is smaller than that of BATSE. Given the
complementary, yet different, characteristics of the Fermi GBM instrument to
BATSE, Swift, and Beppo-SAX, the application of
these statistical clustering
techniques to explore GRB classification 
is the subject of forthcoming studies.  

\section{Summary and outlook}

GRB~170817A/GW170817 has been 
unambiguously identified as resulting from merging neutron stars.
However, GRB~170817A's soft spectrum, intermediate duration,
and unexpectedly faint luminosity 
do not appear to agree with the standard model of Short GRBs.

Over the years many references to GRB classification have unfortunately
devolved into an oversimplified and non-rigorous treatment
based on a theoretical preference for only two GRB
classes (Long and Short) separated
using instrumental-dependent rules deduced from data provided by one
de-orbited instrument (BATSE). In order to resolve
the ambiguity of GRB~170817A's class membership, we
classified GRBs 
by applying statistical clustering methods to
bursts observed by Fermi's GBM,
the same instrument that detected GRB~170817A. 

The classification scheme applied to 1298 Fermi GRBs using 
duration $\mathrm{T}_{90}$ and spectral hardness data.
The choice of the classification parameters and
the assumptions about how GRBs cluster in this parameter space
leads to three classes, which are easily identified as Long, Short, and
Intermediate ones. GRB~170817A/ GW170817 
is most probably identified as an Intermediate burst. 

We conclude that GRB~170817A represents an
Intermediate GRB resulting from a neutron star-neutron star merger.
This is inconsistent with the standard model:
either Intermediate GRBs must be physically very different from Long
GRBs even though their observational properties overlap, or Intermediate GRBs
must be observationally very different from Short GRBs even if they
to originate from similar progenitors.

\cite{2017arXiv171007987B} have recently explored different emission mechanisms to explain the 
unusually weak prompt emission of GRB~170817A, assuming it to be a Short GRB.
They find that synchrotron self-Compton emission from a structured jet might explain
the burst's soft and low-luminosity characteristics. 
If true, then this explanation would indicate that the mechanism producing this kind of merger 
differs from previously accepted mechanisms for Short GRBs. Accordingly,  
\cite{2017arXiv171007987B} suggested that GRB~170817A was a member of a new Short GRB sub-class.
Here we show that their proposed subclass is more likely to be the Intermediate one, 
as earlier discovered by \citet{hor98} and \citet{muk98}.

Relativistic 2D and 3D MHD numerical simulation models have also been developed to 
explain the low-luminosity, 2s duration, and hard-to-soft prompt spectrum of 
GRB 170817A  \citep{kasliwal2017, gottlieb2017, bromberg2017}. Several of these 
models involve a mildly relativistic shock breakout resulting from an asymmetric 
jet interacting with the previously expelled merger ejecta. The low luminosity, 
hard-to-soft prompt $\gamma-$ emission is explained by the jet breakout emanating 
from a wide-angle, mildly-relativistic cocoon.  The models commonly assume that a 
special model is needed to explain the low luminosity and possibly multicomponent 
hard-to-soft prompt spectrum, and that these characteristics are somehow linked.  
However, hard-to-soft multicomponent prompt spectral evolution is a ubiquitous 
feature not only of all Short GRB pulses \cite{hak18} but of all GRB pulses 
\citep{hak14,hak15}, regardless of duration and including Intermediate GRBs. 
MHD models may thus need to be generalized to explain why all GRBs exhibit this 
behavior, rather than attributing it to specific characteristics of an individual GRB.

Other authors also favor the idea that GRB 170817A's low luminosity and soft spectra 
are produced by a structured jet. \cite{meng2018} demonstrate that the soft spectrum of 
GRB 170817A is consistent both with a two-component model in which the spectrum is 
softened by photospheric emission produced in the structured jet and with synchrotron 
emission produced in an optically-thin region. From afterglow observations, 
\cite{margutti2018} find support for the different cases of a mildly relativistic 
spherical ejecta and a structured jet viewed off-axis. \cite{troja2018} use broadband 
observations to demonstrate that Gaussian jet and re-energized cocoon models are favored 
over models involving homogeneous jets, power-law jets, and simple spherical cocoons. 
\cite{lazzati2018} combine numerical simulations with afterglow observations to 
demonstrate that the observed prompt and afterglow characteristics of GRB 170817A 
can be explained by either a structured jet afterglow or a radially stratified spherical 
fireball. \cite{zhang2017} suggest that structured jets might also explain other faint, 
soft GRBs found in the Fermi data archive. This suggestion provides implicit support that 
structured jets produced in some merging neutron star systems might be related to an 
Intermediate GRB class.

More observations of this type of system are clearly needed,
especially by instruments having larger surface areas and greater sensitivity
at lower energies (e.g., Swift).
Further studies will investigate how many such events we may expect to detect.

\begin{acknowledgements} 
We thank the support of the Fermi team, especially Drs. Bhat and Veres, for providing 
the hardness data.
This research was supported by OTKA grant NN111016 and by NASA EPSCoR grant NNX13AD28A. 
Support for \'UNKP-17-3 New National Excellence Program of the Ministry of Human 
Capacities is appreciated (I.R.).

\end{acknowledgements}

\nocite{*}

\end{document}